\documentstyle[12pt,aasms4,psfig]{article}

\begin{document}

\vspace*{-1cm}
\title{DETECTION OF A SUPER STAR CLUSTER AS THE IONIZING SOURCE IN THE 
       LOW LUMINOSITY AGN ~~~~~ NGC 4303}

          \vspace{0.4in}

          \author{Luis Colina\altaffilmark{1},
          Rosa Gonzalez Delgado\altaffilmark{2}, 
          J. Miguel Mas-Hesse\altaffilmark{3}, 
          Claus Leitherer\altaffilmark{4}}

          \vspace{0.15in}

          \scriptsize
          \affil{(1) \em Instituto de Estructura de la Materia (CSIC), 
           Serrano 121, 28006 Madrid, Spain (colina@isis.iem.csic.es)}
          \affil{(2) \em Instituto de Astrof\'{\i}sica de Andalucia (CSIC),
            P.O. Box 3004, 18080 Granada, Spain (rosa@iaa.es)}
          \affil{(3) \em Laboratorio de Astrof\'{\i}sica Espacial y F\'{\i}sica
           Fundamental, INTA, P.O. Box 50727, 28080 Madrid, Spain (mm@laeff.esa.es)}
          \affil{(4) \em Space Telescope Institute, 3700 San Martin Drive,
           Baltimore, MD 21218, USA (leitherer@stsci.edu)}

          \vspace{0.75in}
          \footnote{Based on observations with the NASA/ESA {\it Hubble Space
           Telescope}, obtained at the Space Telescope Science Institute, which
           is operated by the Association of universities for Research in 
           Astronomy, Inc., under NASA contract NAS 5-26555. Based on observations
           with the William Herschel Telescope operated on the island of La Palma
           by the ING in the Spanish Observatorio del Roque de los Muchachos of
           the Instituto de Astrof\'{\i}sica de Canarias.}

          \normalsize

          \begin{abstract}
 
           Hubble Space Telescope (HST) ultraviolet STIS imaging and spectroscopy of the 
           low luminosity AGN (LLAGN) NGC 4303 have identified the previously detected 
           UV-bright nucleus of this galaxy, as a compact, massive and
           luminous stellar cluster. The cluster with a size (FWHM) of 3.1 pc, 
           and an ultraviolet luminosity log L$_{1500\AA}$(erg s$^{-1}$ \AA$^{-1}$)= 38.33  
           is identified as a nuclear super star cluster (SSC) like those
           detected in the circumnuclear regions of spirals and starburst galaxies. 
           The UV spectrum showing the characteristic broad P Cygni lines 
           produced by the winds of massive young stars, is best fitted by the spectral 
           energy distribution of a massive cluster of 10$^{5}$ M$_{\odot}$ 
           (for a Salpeter IMF law with lower-mass cutoff of 1 M$_{\odot}$) generated 
           in an instantaneous burst 4 Myr ago. The ionizing energy produced by this cluster 
           exceeds the flux needed to explain the nuclear H$\alpha$ 
           luminosity. No evidence 
           for an additional non-thermal ionizing source associated with an accreting 
           black hole is detected in the ultraviolet. These new HST/STIS results show 
           unambiguously the presence of a compact, super star cluster in the nucleus 
           of a low luminosity AGN, that is also its dominant
           ionizing source.  We hypothesize that at least some LLAGNs in spirals could 
           be understood as the result of the combined
           ionizing radiation emitted by an evolving SSC (i.e. determined by the mass and age) 
           and a black hole (BH) accreting with low radiative efficiency (i.e. radiating at 
           low sub-Eddington luminosities), coexisting in the inner few parsecs region.           
           Complementary multifrequency studies give the first hints of the very complex 
           structure of the central 10 pc of NGC 4303 where a young SSC apparently coexists 
           with a low efficiency accreting black hole and with an intermediate/old 
           compact star cluster, and where in addition an evolved 
           starburst could also be present. If structures as such detected in NGC 4303 are
           common in the nuclei of spirals, the modeling of the different stellar components, 
           and their contribution to the dynamical mass, has to be established accurately before 
           deriving any firm conclusion about the mass of central black holes of few 
           to several million solar masses.

          \end{abstract}

\keywords { --- galaxies: active --- galaxies: nuclei --- galaxies: starburst --- 
galaxies: individual (NGC 4303) --- ultraviolet: galaxies}

\section{INTRODUCTION}

Low luminosity Active Galactic Nuclei (LLAGNs) including low-luminosity Seyferts, 
classical LINERs, weak$-$[\ion{O}{1}] LINERs, and LINER/HII transition-like objects are the most 
common type of galaxies showing nuclear activity.   
LINERs alone make up 50\%$-$70\% of AGNs and 20\%$-$30\% of all galaxies in surveys of nearby 
bright galaxies (Ho, Filippenko \& Sargent 1997). It is therefore of fundamental importance
to identify unambiguously the nature of the energy source in LLAGNs, and to quantify the
contribution of stars and accreting black-holes to their energy output. 
 
Recent X-ray observations of LLAGNs confirm that some LINERs
are the low luminosity end of the luminous AGN phenomenon identified in Seyfert 1 and QSOs 
where accreting black-holes are the 
dominant energy source (Ho et al. 2001; Eracleous et al. 2002), while in others the energy 
output seems to be dominated by 
powerful starbursts (Terashima et al. 2000; Eracleous et al. 2002).
HST ultraviolet imaging and spectroscopy has shown the presence of young massive
stars at various scales in AGNs. Seyfert 2 galaxies known to have bright kpc-size  
star-forming rings show that the AGN core is barely detected in the UV, and that the 
massive stars dominate the observed circumnuclear UV emission (Colina et al. 1997a). 
The detection of 
stellar winds and photospheric absorption lines in the ultraviolet spectra of Seyfert 2 
galaxies (Heckman et al. 1997; Gonz\'alez Delgado et al. 1998) has proven unequivocally 
the presence of clusters of massive young stars in the circumnuclear regions of 
these galaxies. These nuclear starbursts are dusty star-forming regions of a few hundred 
parsecs in size and with an average age of 3 to 6 Myrs (Gonz\'alez Delgado et al. 1998).
Optical and UV studies of Seyfert 2 galaxies have detected the presence of young massive
starbursts within 300 pc of the nucleus in 30\% to 50\% of the galaxies investigated
(Cid Fernandes et al. 2001 and references therein).
Finally, the compact ultraviolet sources detected in some UV bright LINERs are 
nuclear star clusters with sizes of about 10-15 pc like in NGC 4569 (Maoz et al. 1998; 
Barth et al. 1998). 

In summary, recent evidence collected mainly with HST and {\it Chandra} indicates the presence 
of massive stellar
clusters in the circumnuclear regions (few to several hundred parsecs) in a large
fraction of LINERs and
Seyfert 2 galaxies, that substantially contribute to their energy output.  However,  
the fundamental question of whether or not massive stellar clusters exist in the central few 
pc region, i.e. nucleus, 
of LLAGNs contributing substantially, or even dominating, 
the energy output, requires detailed investigations in selected nearby 
galaxies like NGC 4303. 

NGC 4303 (M 61) is a barred spiral classified as SAB(rs)bc (de Vaucouleurs et
al. 1991) and located in the Virgo cluster (adopted distance of 16.1 Mpc
hereinafter; see also Colina et al. 1997b).
Multi-wavelength HST images of NGC 4303 have unveiled the presence
of a nuclear stellar bar of 250 pc in size centered on a bright optical and 
near-IR nucleus\footnote{throughout the paper we distinguish between nucleus, nuclear
and circumnuclear regions each indicating different physical sizes of about 0.01, 0.1 and 1 kpc,
respectively}, itself connected with 
a nuclear star-forming spiral of about 250 pc in radius. 
(Colina et al. 1997b; Colina \& Arribas 1999; Colina \& Wada 2000). 
The UV luminosity of the spiral dominates the observed integrated UV output.  
The brightest knots delineating the spiral have observed UV luminosities
 L(2200\AA) $\sim$ 2 $\times$ 10$^{37}$ erg s$^{-1}$ \AA$^{-1}$ similar to that of the
R136 cluster in 30 Doradus (Vacca et al. 1995), and in the
high UV luminosity end of the distribution function of compact stellar
clusters detected in star-forming rings (Maoz et al. 1996).  

The 2D velocity field of the warm ionized gas and cold molecular 
gas shows that the nucleus is also the dynamical center of a disk of 300 pc 
in radius, and in which the star-forming spiral
is embedded (Colina \& Arribas 1999; Schinnerer et al. 2002; Colina et al. 2002, 
unpublished).
 Ground-based spectroscopy reveals that the optical 
emission lines emitted by 
the nuclear region ($\sim$ 100 pc size) have ratios 
in the borderline of Seyfert 2 and LINER nuclei (Colina \& Arribas 1999).
These observations confirm that the bright nucleus of NGC 4303 is 
a low luminosity AGN and is indeed located at the true center of the galaxy.

\section{OBSERVATIONS}

Ultraviolet imaging and spectroscopy of NGC 4303 was obtained using STIS onboard
HST. The ultraviolet image was taken on July 7th, 2000 through filter F25QTZ 
($\lambda_p$= 2364.8\AA) 
for a total integration time of 2080 seconds. Spectroscopy was done on February 12th, 2001
with the 
52\farcs0 $\times$ 0\farcs2 long-slit and G140L grating covering the 1100-1700\AA~
spectral range with a dispersion of 0.6\AA/pix. The roll angle of the telescope
during the observations was such that the slit was oriented along position angle 220
degrees, i.e. along the line connecting the UV-bright nucleus of NGC 4303 with the 
brightest circumnuclear stellar cluster as identified in the previous HST/WFPC2 ultraviolet 
image (cluster G; Colina et al. 1997b). Total integration time for the spectroscopy
was 13442 seconds divided in five individual exposures taken during five contiguous orbits,
and using therefore the same guide stars. The individual imaging and spectroscopic exposures
were dithered and therefore they were calibrated independently with the
standard STIS calibration pipeline using the most updated calibration files. The individual
images and spectra were combined afterwards to generate the final image and 
long-slit spectrum.   

A high signal-to-noise long-slit optical spectrum was obtained with the ISIS double 
spectrograph attached at the William Herschel Telescope during an observatory service
observing run on May 29th, 2001. The 1\farcs0 wide slit was oriented along position angle 
220 degrees on the sky to cover the nucleus and circumnuclear clusters. The total integration 
time was 3600 seconds, split in three equal 
exposures of 1200 seconds, each. The seeing conditions were stable and values of 
1\farcs5 - 2\farcs0 were
measured. The final spectra were obtained combining two 
independent blue (3600$-$6280 \AA) and red (6205$-$6990) spectral ranges taken 
simultaneously with the two-arm spectrograh. The spectral dispersion was 0.8\AA/pix and 
0.9\AA/pix for the red and blue 
ranges of the spectrum, respectively. Calibration was done following standard long-slit
spectroscopic procedures. These data are used here only to measure the H$\alpha$ nuclear flux,
and to obtain a new determination of the optical emission line ratios (see Table 1). 
Although the airmass was in the 1.3 to 1.55 range during the observations,
atmospheric differential refraction has no effect on the emission line ratios since the slit
was positioned close to the parallactic angle (off by up to 10 degrees).   

\section{RESULTS}

The high-resolution (0\farcs025 per pixel) deep STIS ultraviolet image (Figure 1) shows in 
much more detail the structure already detected in a previous lower resolution (0\farcs1 
per pixel) WFPC2 image (Colina et al. 1997b). Each of the previously unresolved star-forming 
knots located
in the circumnuclear spiral structure, breaks now into several smaller, fainter knots separated
by distances of less than 0\farcs2 (i.e $\leq$ 15 pc; see Figure 1, and WFPC2 UV image in
Colina 1997b for a comparison). However, the 
UV-bright nucleus, unresolved in our previous WFCP2 image, remains as a compact
source. Although
a full analysis of the STIS image is beyond the scope of this paper and the results 
will be published elsewhere (Colina et al. in preparation), it is relevant for the purpose 
of this paper to establish the size of the compact UV-bright nucleus. 
The STIS PSF for the filter F25QTZ was modeled using Tiny Tim version 6.0 (Krist \& Hook 2001)
and assuming a 
flat spectrum (F$_{\lambda}$= constant) source. Encircled energy measurements and 
two-dimensional gaussian fits to the core of the Tiny Tim PSF light profile were performed, 
and subsequently compared with the results obtained for the observed light  
distribution of the UV-bright nucleus. The results of this analysis allows one 
to conclude that the nucleus of NGC 4303 is resolved and has a size (FWHM)  
of 0\farcs040 ($\pm$ 0\farcs005) in the UV, equivalent to 3.1 pc at the 
assumed distance of NGC 4303. In addition, simulations of extended sources of various sizes 
convolved with the modeled Tiny Tim PSF were also performed. For these simulations, the modeled 
PSF was computed considering a subsampling factor of 5, 
i.e. 0\farcs005 per pixel. Extended sources were simulated as regions of constant surface 
brightness with diameters ranging from 0\farcs015 to 0\farcs055. The extended sources 
were convolved with the subsampled PSF and the resulting images were rebinned by a factor 
5 to generate the final simulated images with a pixel size of 0\farcs025, i.e. the pixel scale
of STIS detectors in the ultraviolet. As before, encircled 
energy measurements and two-dimensional gaussian fits were performed on the 
rebinned images. The size of the nucleus obtained in this way is consistent with the value
derived from the previous analysis, and corresponds to that of 
an extended region of 0\farcs045 in diameter.   

The STIS UV long-slit spectrum along PA 220 shows three independent spectra associated
with the nucleus, the brightest circumnuclear cluster (cluster G after 
Colina 1997b), and a fainter cluster (cluster L, not previously identified). 
The one-dimensional spectra fro the nucleus, cluster G and cluster L were extracted from the
long-slit spectrum using apertures of 0\farcs9, 0\farcs8 and 0\farcs5, respectively. 
The three spectra are almost indistinguishable (see Figure 2) showing all
of them very prominently the characteristic broad P Cygni lines (\ion{N}{5} 1240\AA, 
\ion{Si}{4} 1400\AA, \ion{C}{4} 1550\AA) produced by the winds of massive young stars. 
In addition, the 
three spectra show a narrow Ly$\alpha$ line in emission but no trace of other strong
emission lines like \ion{Si}{4} 1400\AA, \ion{C}{4} 1550\AA, and \ion{He}{2} 1640\AA\ typical 
of luminous 
Seyfert 2 galaxies like NGC 1068 (Crenshaw \& Kraemer 2000),
 or gas heated by fast shocks (Allen et al. 1998). 

In order to establish the age and mass of the stellar clusters, as well as the internal
absorption towards them, a detailed comparison of the observed and synthetic 
ultraviolet spectra has been performed assuming a given extinction law. Differences between 
Calzetti's law, derived from ultraviolet observations of starburst galaxies (Calzetti et 
al. 2000), and LMC extinction law can not be appreciate due to the small
wavelength range covered by the STIS spectra. However, extinction in star-forming regions 
seems to follow an LMC or SMC-like law, i.e. with a rather weak bump at 2200 \AA, 
independently of the metallicity
of the region (Mas-Hesse \& Kunth 1999). These authors concluded that the high ionizing flux
produced by the young massive stars would destroy the graphite grains, which are responsible for
the shape of the Galactic extinction law, leaving mostly silicates, whose extinction
properties are closer to the shape of the LMC-SMC laws. Therefore, a LMC extinction law is used
throughout the following analysis. The extinction values using Calzetti's law are also given
in Table 1 for completeness. 
The ages of the clusters have been obtained by comparing the 
observed profile of the P Cygni lines with the model predictions for solar metallicity stars.
Synthetic spectra with metallicities lower than solar (obtained using a mix of LMC and
SMC stellar libraries)
do not reproduce the intense P Cygni lines detected in NGC 4303. Moreover, instantaneous bursts
are preferred over continuous star formation because the \ion{Si}{4} 1400\AA~ profile is better 
fitted by instantaneous bursts of a given age. Finally, once the age of the clusters and 
stellar upper mass of the initial mass function (IMF) are fixed, the mass and internal 
extinction of the clusters are derived 
by comparing the slopes of the observed and synthetic (stellar plus extinction) continua 
over the entire ultraviolet 
spectral range available to us.    

Following the methodology outlined above, the UV spectrum of the nucleus with its prominent 
broad P Cygni lines is best fitted 
with the  synthetic spectrum of an unobscured (E(B$-$V)= 0.07 for an LMC extinction law), 
massive 1 $\times$ 
10$^{5}$ M$_{\odot}$, 4 Myr old instantaneous starburst, with a Salpeter IMF in the 
1 to 100 M$_{\odot}$ stellar mass range (see Figure 3 and Table 2). 
The circumnuclear clusters are similarly best fitted by unobscured, less
massive ($\sim$ 10$^{4}$ M$_{\odot}$), younger (3 to 3.5 Myrs) clusters (see
Table 2 for detailed parameters). The best fit for cluster G requires an IMF characterized
by a slope of 1.5, i.e. flatter than Salpeter, indicating the need for a large fraction of
very massive stars to explain the UV spectral features. The need for a flat IMF does not
necessarily imply an IMF different than Salpeter but could be 
understood if cluster G were extended and therefore, due to mass segregation within the 
cluster, we could be biased towards the more massive stars
since the width of the slit is only of 0\farcs2 (i.e. 15.6 pc) and its
orientation was selected along the line connecting the UV-bright nucleus
and the UV-peak emission of cluster G.

Although the UV spectra of the nucleus and circumnuclear stellar clusters are almost
indistinguishable, the optical emission line ratios derived from the WHT narrow long-slit 
spectrum show that the excitation conditions in the nucleus
are clearly different from those present in the circumnuclear stellar clusters (see Figure 4
and Table 1).
While the circumnuclear clusters show the typical emission line ratios of compact HII
regions, the spectrum of the nucleus (1\farcs0 $\times$ 1\farcs0) is classified as a LINER 
or, due to the weakness of the [\ion{O}{1}] 6300\AA\
line, as a weak-[\ion{O}{1}] LINER (Figure 4). However, our previous two-dimensional 
integral-field spectra 
identified the nucleus as a low-luminosity Seyfert 2/LINER borderline AGN (Colina \& 
Arribas 1999). Discrepancies between these two classifications illustrate the difficulties 
in classifying LLAGNs, mostly due to the 
intrinsic uncertainties in ground-based measurements in general, i.e. positioning of a narrow 
slit (1\farcs0), corrections in the Balmer emission lines due to stellar absorptions, 
structure and size of extended ionized gas, measurement of weak emission lines against the bright
continuum emitted by the bulge population, etc. For NGC 4303, the structure of the ionized gas
and the contribution of the H$\beta$ line in absorption could be relevant factors working on 
opposite directions when computing the \ion{O}{3}/H$\beta$ ratio. The surface 
brightness of the
H$\beta$ and [\ion{O}{3}] 5007\AA~ lines along PA 220 indicates that the high-excitation 
[\ion{O}{3}] emitting gas is more centrally concentrated than the low-excitation H$\beta$ 
emitting gas, as traced by the steep gradient in the [\ion{O}{3}] light distribution with 
the available 0\farcs3 pixel resolution. This could have an important effect decreasing the
[\ion{O}{3}]/H$\beta$ ratio when obtaining the integrated spectrum for a region of 1\farcs0 
to 1\farcs5 across. Therefore, this suggest that the [\ion{O}{3}]/H$\beta$ ratio would 
increase if an spectrum sampling the central 0\farcs1 were available. On the other hand, 
the nuclear optical continuum of NGC 4303 redwards 
of 4800\AA~ can be fitted reasonably well with a 1 Gyr stellar population. The core of the
corresponding stellar H$\beta$ absorption line with an equivalent width of 1.8\AA~ will 
decrease the flux of the H$\beta$ nebular line by a factor of about two. If this 
correction were applied to the observed [\ion{O}{3}]/H$\beta$ ratio, its value would drop by 
a factor 2 with respect 
to the value given in table 1. Therefore although the most recent ground-based spectra favors
the classification of the LLAGN nucleus of NGC 4303 as a [\ion{O}{1}]-weak LINER,
the final classification will not be obtained until the new scheduled 
HST/STIS spectroscopic data, isolating the emission from the nucleus with a slit width
of 0\farcs2 and covering the entire 1200\AA\ $-$ 9000\AA\ range, are taken.  

Finally, the internal extinction towards the nucleus and circumnuclear stellar
clusters is very low with E(B$-$V) values of less or equal than 0.1 as derived from 
the UV-shape of the spectrum (see Table 1 for specific values). These values agree with
those derived from the H$\alpha$/H$\beta$ line ratios after the fluxes of the nebular 
Balmer lines are corrected by an stellar absorption of 1.8\AA (see above).

\section{DISCUSSION}

\subsection{NGC 4303 nucleus: A LLAGN powered by  super star cluster}

The nuclear cluster characterized by its size of 3.1 pc, its mass of 10$^5$ M$\_{\odot}$,
and its ultraviolet luminosity log L$_{1500\AA}$(erg s$^{-1}$ \AA$^{-1}$)= 38.33, or
log L$_{1500\AA}$(erg s$^{-1}$)= 41.51 assuming $\nu$ $\times$ f$_{\nu}$ 
$\times$ 4$\pi$D$^2$ as the
monochromatic luminosity, belongs
to the class of luminous super star clusters (SSCs) found at the heart of spirals 
(Carollo et al. 1997), and in nuclear starburst galaxies (Meurer et al. 1995). At the derived
age of 4 Myrs, the SSC in the nucleus of NGC 4303 is extremely luminous with a bolometric 
luminosity of  
about 10$^8$ L$_{\sun}$, and an ionizing flux capable of producing an H$\alpha$ luminosity 
of up to about 1.7 $\times$ 10$^{39}$ erg s$^{-1}$, if all ionizing photons are absorbed by
the surrounding interstellar medium. 

The H$\alpha$ flux in the 1\farcs0 $\times$ 1\farcs0 nuclear region along PA220 
corresponds to 
a luminosity of 1.2 $\times$ 10$^{39}$ erg s$^{-1}$, after correcting for  
an stellar absorption of 1.8\AA~ (equivalent width), and after slit/seeing aperture 
correction effects have been taken into
account. The ratio of the predicted H$\alpha$ flux emitted by the SSC to the measured nuclear
H$\alpha$ flux is 1.4.  Therefore no additional ionizing source other than the SSC itself is
required to explain the ionized gas luminosity.

\subsection{NGC 4303 nucleus: coexisting star clusters and AGN}

As mentioned above, no evidence for a second energy source is present in
the ultraviolet spectrum of the nucleus. However, its absolute optical
magnitude M$_{F606W}$= $-$14.2 and very red nuclear colors
m$_{F606W}-$m$_{F160W}$= $+$3.5 obtained from HST 0\farcs2 radius
aperture measurements with filters WFPC2/F606W and NIC2/F160W\footnote{Although filters
F606W and F160W have a broadband profile different from the standard ground-based V and 
H filters, their effective wavelengths are very similar and consequently we adopt here 
the same names for simplicity.} (see also
Colina \& Wada 2000), can not be explained by the emission due to the
unobscured UV-bright SSC.  According to the STARBURST99 models (Leitherer
et al. 1999), the 4 Myrs, 10$^5$ M$_{\sun}$ UV-bright cluster should have
an absolute visual magnitude of about $-$13, i.e. three times fainter than
measured, and an optical $-$ near-infrared color V$-$H $\sim +$0.2, i.e. much
bluer than measured. Optical emission lines could affect the measured F606W
magnitude, and therefore the observed V$-$H color could not represent the
intrinsic continuum value, that would be even redder than observed.  The WFPC2/F606W
filter is an extreme broad-band filter with a bandpass of almost 1600\AA~
that includes all the optical emission lines in the 4800\AA~ to 7000\AA~
spectral range. The predicted equivalent widths of the Balmer lines
produced by a 4 Myr cluster are about 100\AA~ and 400\AA~ for H$\beta$ and
H$\alpha$, respectively (Leitherer et al. 1999).  However, the ionized
region centered on the nucleus is extended over about 1\farcs5, as
traced by the angular size of the Ly$\alpha$ emission line region in our STIS long-slit
spectrum. Moreover, the combined equivalent width of the nuclear
[NII]+H$\alpha$ line complex, as measured in our WHT spectrum of the nuclear region,
corresponds to about 20\AA~ indicating dilution by the bulge starlight contribution,
that dominates the optical continuum in our ground-based spectrum. 
In summary, the contribution of the optical emission lines to the
measured nuclear F606W flux is not expected to be more than a few percent.

The nuclear m$_{F606W}-$m$_{F160W}$ color is also redder than the average
bulge's color in face-on spirals (V$-$H= 2.71 $\pm$ 0.33; de Jong, \& van
der Kruit 1994) and therefore, the presence of an additional red and
luminous source has to be invoked. This additional source could either be
an intermediate/old stellar population associated with the bulge, 
an evolved starburst
dominated by red supergiants (about 10 Myr), or an accreting black hole. In
the following these alternatives are discussed.

The first possibility would be that the bright near-infrared source traces
the presence of an intermediate/old nuclear star cluster. Optical HST imaging has shown
that many nearby spirals harbor nuclear star clusters, with a small
fraction (about 5\%) being unresolved (Carollo et al. 1997). Our long-slit optical
continuum integrated over 1\farcs0 $\times$ 1\farcs0 seems indeed to be
dominated by an evolved stellar population around 1-5 Gyr old, with a total
initial mass around 10$^8$ M$_\odot$ (Salpeter IMF normalized between 1 and
100 M$_\odot$). Even allowing for some reddening (E(B-V) of 0.4 and
0.1 for both ages, respectively), the V$-$H color of this population
wouldn't be redder than around 2.5, not being able to explain
completely the observed infrared luminosity.  

The excess H-band luminosity could also originate, at least partially, from a
population of red supergiant stars formed in a previous starburst episode
around 10 Myr ago (Cervi\~no \& Mas-Hesse 1994; Leitherer et al. 1999), so that a 
two-stage starburst
consisting of 4 + 10 Myr old stars within a few pc could coexist in the
nucleus of NGC 4303.  This scenario of a two-stage starburst is reminiscent of
the star formation observed in the central 0.5 pc of the Milky Way and in
NGC 1569, where a recent 4-8 Myr starburst, fully accounting for the
ionizing and bolometric luminosities, coexist with an older cluster (Krabbe
et al. 1995; Gonzalez Delgado et al. 1997).  If a 10 Myr cluster were the
dominating source of the luminous H-band point source (M$_{F160W}$= $-$17.7)
in NGC 4303, it would have had a mass of about 2 $\times$ 10$^6$
M$_{\odot}$.  If we include the H-band luminosity associated with the very 
intermediate/old
($>$ 1 Gyr) stellar population, a starburst of about 5 $\times$
10$^5$ M$_{\odot}$ would be required. In either case, such a massive,
unobscured cluster would produce an ultraviolet flux in excess of what is
observed, and would produce a much bluer optical continuum than
observed.  Therefore, if the 10 Myr old starburst were present, it should be very
significantly obscured both in the optical and ultraviolet ranges, with its
associated light emerging only in the infrared, if at all.

Finally, the nuclear, luminous near-infrared source could alternatively
indicate the presence of an AGN. The radiation emitted by the accreting black hole
would dominate the energy output in the near and mid-infrared, as in
most Seyfert 2 galaxies (Alonso-Herrero et al. 2001). Recent
near-infrared imaging with HST has shown that all surveyed Seyfert 1 and
50\% of the Seyfert 2 galaxies contain a luminous unresolved continuum
source at 1.6 $\mu$m (Quillen et al. 2001).  The H-band luminosity of the
NGC 4303 nucleus (M$_{F160W}$= $-$17.7) agrees with the average value obtained by
Quillen and collaborators for the subsample of Seyfert 2 galaxies with
unresolved nuclear sources. There are additional
independent indications that an accreting black hole exist in the
nucleus of NGC 4303. The ground-based optical spectrum of the nuclear region 
($\sim$ 100 pc) have line ratios anywhere between weak$-$[\ion{O}{1}] LINERs and 
Seyfert 2 galaxies (see Table 1). However, 
the strongest evidence for an accreting black hole might come from recent
{\it Chandra} images (Jim\'enez-Bail\'on et al.  2002, in preparation)
that reveal the presence of an unresolved hard X-ray 
(2$-$10 keV) power-law source astrometrically coincident with the UV-bright nucleus,
and similar in luminosity to other LLAGNs recently studied with {\it Chandra} 
(Ho et al. 2001). 
Following recent determinations of the black hole mass to velocity dispersion 
relationship (Ferrarese \& Merritt 2000; Tremaine et al. 2002), the central velocity 
dispersion of 74 km s$^{-1}$ (Heraudeau \& Simien 1998) implies a mass of 1.2 
to 2.5 $\times$ 10$^6$ M$_{\odot}$ for the central black hole in NGC 4303.
Given this mass, the X-ray accreting
black hole would radiate very unefficiently, at extremely low
sub-Eddington luminosities as in other LLAGNs (Terashima et al. 2000).

In summary, the high spatial resolution multifrequency studies done so far,
give the first hints of the very complex structure of the central 10 pc of NGC 4303
where a young, luminous SSC apparently coexists with a low efficiency
accreting black hole and with an intermediate/old star cluster. The young SSC is the
dominant ionizing source, the accreting black hole is a minor contributor to the overall
ionization and the old cluster contributes substantially to the optical and near-ir 
flux. Some additional red supergiant stars associated with an evolved starburst could
also contribute to the near-infrared continuum.

\subsection{Implications for LLAGNs: the SSC$-$AGN connection}

Low-luminosity AGNs as the one identified in the nucleus of NGC 4303,
make up the vast majority of the AGN
population (Ho, Filippenko \& Sargent 1997).   
The empirical evidence obtained so far indicates that a powerful SSC seems to coexists 
with an accreting black hole within the central 3 pc of NGC 4303.
SSCs as the one detected in NGC 4303 are a common phenomenon in the nuclear regions 
of early and late-type spirals (Carollo et al. 1997; Carollo, Stiavelli \& Mack 1998; 
Carollo et al. 2002; Boeker et al. 2002), and in galaxies with nuclear 
starbursts and circumnuclear star-forming rings (Meurer et al. 1995; Maoz et al. 1996). 
Therefore SSCs are a natural consequence of the star formation processes in the nuclear 
regions of spirals. On the other hand, the tightness of the black hole mass and stellar 
velocity dispersion relation (Ferrarese \& Merritt 2000 and references; Tremaine et al. 2002) 
implies a link between massive 
black holes (MBH) and bulge formation in galaxies, and therefore nuclear MBHs 
should also be a natural consequence of the physical processes that formed 
present-day galaxies. So, it is reasonable to hypothesize that, as detected in the 
LLAGN NGC 4303, 
powerful SSCs could coexist with MBHs in the nucleus, i.e. inner few parsecs, of a large
fraction of spirals. 
Under this hypothesis, at least some types of LLAGNs
could be understood primarily as a consequence on one hand of the 
age and mass of the SSC, and on the other hand of the accretion 
rate and mass of the BH. Massive SSCs with masses of 
10$^5$ to 10$^6$ M$_{\odot}$ would have peak bolometric luminosities 
of 0.2 to 2 $\times$ 10$^9$ L$_{\sun}$ 
at an age of 3 Myrs. The associated, unobscured, H$\alpha$ luminosities produced in the
ionized interstellar medium surrounding the SSC would be in the 0.14 
to 1.4 $\times$ 10$^{40}$ erg s$^{-1}$ range. On the other hand, low mass black holes such
as the one in NGC 4303 (see $\S$4.2) 
surrounded by accretion disks radiating at extremely 
low sub-Eddington luminosities ($\sim$ 10$^{-4} - 10^{-5}$ L$_E$; Terashima et al. 2000), 
would emit bolometric luminosities of 
10$^6$ to 10$^7$ L$_{\odot}$, i.e. a factor of hundred less than a young SSC. 

Therefore some types of LLAGNs could be understood as the result of 
the combined ionizing radiation emitted by an evolving super star cluster (i.e. 
determined by the mass and age) and an accreting black hole (i.e. radiating at 
sub-Eddington luminosities), coexisting in the inner few parsecs region. The SSC would
have an ionizing spectral energy distribution peaking in the soft X-rays/far-UV while
the accreting black hole would have a harder radiation field with substantial flux beyond
1 keV.  The nucleus of NGC 4303 with a central young (3-4 Myr old) SSC 
dominating its ionizing output in the 
inner few parsecs region, could be a prototype of the class of  
weak$-$[\ion{O}{1}] LINERs. The LINER/HII nucleus in NGC 4569 could be another example
of SSC-dominated LLAGN (Maoz et al. 1998; Barth \& Shields 2000; Grabel \& Bruhweiler 2002).
On the other hand, LLAGNs like classical LINERs, or even low luminosity 
Seyfert 2 nuclei, could still host an evolved, i.e. 10 Myr or older, nuclear stellar cluster 
that will have a minor contribution to the ionizing luminosity, dominated therefore by an 
accreting  black hole.

\subsection{Implications for LLAGNs: measuring the low-end of the black hole mass function}

The complementary multifrequency study of the nucleus of NGC 4303 has shown that 
stellar clusters of different ages seem to coexist with a black hole in the central few parsecs
region of this LLAGN. Even the best 0\farcs1 spatial resolution available with HST 
STIS spectrograph represents a linear resolution of 10 pc at a distance of 20 Mpc, not enough 
to resolve the sphere of influence of nuclear black holes with masses of less than 10$^7$ 
M$_{\odot}$. Therefore, the mass contribution of massive 
young SSCs as the one detected in NGC 4303, and of massive, compact, intermediate/old 
nuclear stellar clusters in spirals with central low mass black holes (masses of a few million
solar masses as in the Milky Way, or as the estimated in NGC 4303, see section $\S$4.2) could 
not be negligible and has to be taken into account. Black hole mass measurements are generally 
done under the assumption that the mass to light ratio of the stars is the same for all 
spirals, and spatially constant over the region used for the measurement (Sarzi et al. 2001;
Sarzi et al. 2002). 
This might not be a bad assumption for ellipticals, but spirals could have large differences, 
in M/L, or even gradients within same galaxy, if there are young nuclear clusters of different 
ages. NGC 4303 is
a clear example of a spiral where the assumption of constant M/L would not be valid. Therefore, 
kinematical studies based on the analysis of optical emission lines alone can 
not provide the mass contribution of nuclear clusters, and therefore additional detailed 
multifrequency modeling and spectroscopy with HST would be required before deriving any 
reliable mass for central black holes with masses of a few to several million solar masses.

\section{Summary}

           The main results presented in this paper can be summarized as follows:

   (1)     New HST ultraviolet STIS spectrum of the
           nucleus of the galaxy NGC 4303, classified as a LLAGN in the optical, shows 
           unambiguously the presence of broad and intense P Cygni lines characteristic 
           of young, massive stars and do not show any evidence of the strong UV emission 
           lines characteristic of classical AGNs like NGC 1068. 

   (2)     The ultraviolet properties of the nuclear cluster correspond to that of a 
           compact (3.1 pc), young (4 Myrs), massive (10$^{5}$ M$_{\odot}$),  and 
           luminous (log L$_{1500\AA}$(erg s$^{-1}$ \AA$^{-1}$)= 38.33) star cluster. These
           properties are characteristic of the so called super star clusters (SSCs) 
           commonly detected in the (circum-)nuclear regions of spirals and starburst galaxies.  
 
   (3)     The SSC is the dominant ionizing source in the nucleus of NGC 4303.
           The ionizing energy produced by this cluster exceeds 
           the flux needed to explain the extinction corrected nuclear H$\alpha$ luminosity,
           and therefore an additional non-thermal ionizing source is not required. 

   (4)     New ground-based optical spectrum of the nuclear region (100 pc $\times$ 100 pc)
           favors the classification
           of the LLAGN as a weak$-$[\ion{O}{1}] LINER although previous classification
           identified it as a LINER/Seyfert 2 borderline LLAGN. Scheduled HST observations
           will establish the final classification.

   (5)     Circumnuclear stellar clusters at distances of 170 to 230 pc from the nucleus
           have similar ultraviolet spectra showing the broad and intense P Cygni lines 
           characteristic of young, massive stars. These clusters are however a bit younger 
           (3 to 3.5 Myr) and less massive (7$-$8 $\times$ 10$^3$ M$_{\odot}$)
           than the SSC detected in the nucleus.    
         
   (6)     Additional multifrequency studies give the first hints of the very complex 
           structure of the nucleus, i.e. region of a few pc in radius, of NGC 4303 where 
           the SSC apparently 
           coexist with a low efficiency accreting black hole and with an intermediate/old 
           ($>$ 1 Gyr) compact star cluster, and where in addition an evolved ($\sim$ 10 Myr) 
           starburst could also be present.

    If the structure detected in the nucleus of NGC 4303 were common in
    spirals, there are two important implications that deserve further investigation:

    (i)    Some types of LLAGNs should be understood as the result of 
           the combined ionizing radiation emitted by an evolving super star cluster (i.e. 
           determined by the mass and age) and an accreting black hole (i.e. radiating at 
           sub-Eddington luminosities), coexisting in the inner few parsecs region.
           Under this scheme, the ionization in LLAGNs classified as LINER/HII nuclei or 
           weak$-$[\ion{O}{1}] LINERs would be dominated by a young (3-4 Myrs) super star 
           cluster. Classical LINERs and low-luminosity Seyfert 2 could still host an 
           older ($\geq$ 10 Myrs) cluster but the ionizing continuum would be dominated by
           an accreting black hole.
                       
    (2i)   For spirals containing low mass (i.e. less than few to several 10$^6$ M$_{\odot}$)
           nuclear black holes, the contribution of massive nuclear clusters to the dynamical 
           mass within the inner 10 pc region has to be established 
           accurately before deriving any firm conclusion about the mass of the black hole.

\acknowledgments

     LC, RGD, and JMMH acknowledges support by the Spanish Ministry of Science and 
     Technology (MCYT) through grants PB98-0340-C02, AYA-2001-3939-C03-01 and 
     AYA-2001-3939-C03-02. The authors want to thank the anonymous referee for his/her
     excellent report that has largely improved the clarity and content of the paper.
     This research 
     was supported by grant GO-08665.01-A from the Space Telescope Science Institute.
     Authors thank the staff at El Roque de los Muchachos Observatory that
     obtained the optical spectra during the course of service observations with the
     William Herschel Telescope.

\newpage

     \scriptsize

     \begin{deluxetable}{cccccccc}
     \tablewidth{41pc}
     \tablecaption{Ultraviolet and optical properties of the nucleus and 
       circumnuclear clusters}
     \tablehead{
     Region& F$_{UV}$(1500\AA)\tablenotemark{a} & E(B$-$V)\tablenotemark{b} & 
     logL$_{UV}$(1500\AA)\tablenotemark{c} &
     log([\ion{O}{3}]/H$\beta$)\tablenotemark{d} & log([\ion{O}{1}]/H$\alpha$)\tablenotemark{d} 
      & 
     log([\ion{N}{2}]/H$\alpha$)\tablenotemark{d} & log([\ion{S}{2}]/H$\alpha$)\tablenotemark{d} 
     \\}
     \startdata
     Nucleus  & 3.40 & 0.07 (0.10) & 38.33 & 0.28 & $-$1.35 & 0.06 & $-$0.32 \nl
              &      & 0.0 &       & 0.50 & $-$0.96 & 0.01 & $-$0.31 \nl
     Cluster G & 0.84 & 0.10 (0.15) & 37.85 & $-$0.37 & -- & $-$0.27 & $-$0.54 \nl
               &      & 0.1     &       & $-$0.72 & -- & $-$0.35 & $-$0.64 \nl
     Cluster L & 0.25 & 0.10 (0.15) & 37.33 & -- & -- & -- & -- \nl
     \tablenotetext{a}{Observed ultraviolet flux, not corrected for extinction, in units
       of 10$^{-15}$ erg s$^{-1}$ cm$^{-2}$ \AA$^{-1}$.}
     \tablenotetext{b}{Internal extinction. First row gives the value as derived from the 
      UV spectra considering an LMC extinction curve. Values in parenthesis were
      derived using Calzetti's law (Calzetti et al. 2000).
      Second row presents the value derived from the H$\alpha$/H$\beta$ ratios.}
     \tablenotetext{c}{Extinction-corrected luminosity using the listed E(B-V) values in
      units of erg s$^{-1}$ \AA$^{-1}$.}
     \tablenotetext{d}{Extinction-corrected emission line ratios. In addition to the 
      hydrogen Balmer lines, the collisionaly-excited lines used are
           [\ion{O}{3}] 5007\AA, [\ion{O}{1}] 6300\AA, [\ion{N}{2}] 6584\AA, and the  
           [\ion{S}{2}] doublet 6717,6731\AA.     
                   First row of values as derived from the long-slit WHT/ISIS spectrum
                   using an aperture of 1$^{''} \times 1^{''}$. 
                   Second row are the values derived from previous
                   integral field spectroscopy (Colina \& Arribas 1999). The spectrum
                   for cluster L, at a
                   distance of 0.$^{''}$6 from cluster G, can not be obtained in
                   ground-based spectroscopic observations.}
     \enddata
     \end{deluxetable}

\scriptsize

     \begin{deluxetable}{ccccccc}
     \tablewidth{37pc}
     \tablecaption{Modeled properties of the nuclear and circumnuclear 
     stellar clusters}
     \tablehead{
     Region & IMF index & Age & Mass & log Q$_{model}$\tablenotemark{a} & 
     log Q$_{H\alpha}$\tablenotemark{b} \\ 
     & & (Myrs) & (10$^{4}$ M$_{\odot}$) & (ph s$^{-1}$) & (ph s$^{-1}$)\\}
     \startdata
     Nucleus  & 2.35 & 4.0 & 10.0 & 51.10 & 50.9 \nl
     Cluster G & 1.50 & 3.0 & 0.7 & 51.04 & 50.4 \nl
     Cluster L & 2.35 & 3.5 & 0.8 & 50.22 & -- \nl
     \tablenotetext{a}{Q$_{model}$ is derived integrating the synthetic spectral energy 
      distribution of the modeled clusters for photons with energies above 13.6 eV}
     \tablenotetext{b}{Derived from the WHT/ISIS H$\alpha$ flux after correction for
      stellar absorption and aperture effects, and assuming Q$_{H\alpha}$ is given as
      7.5 $\times$ 10$^{11}$ $\times$ L(H$\alpha$) ph s$^{-1}$}     
     \enddata
     \end{deluxetable}


\begin{figure} 

\centerline{\psfig{figure=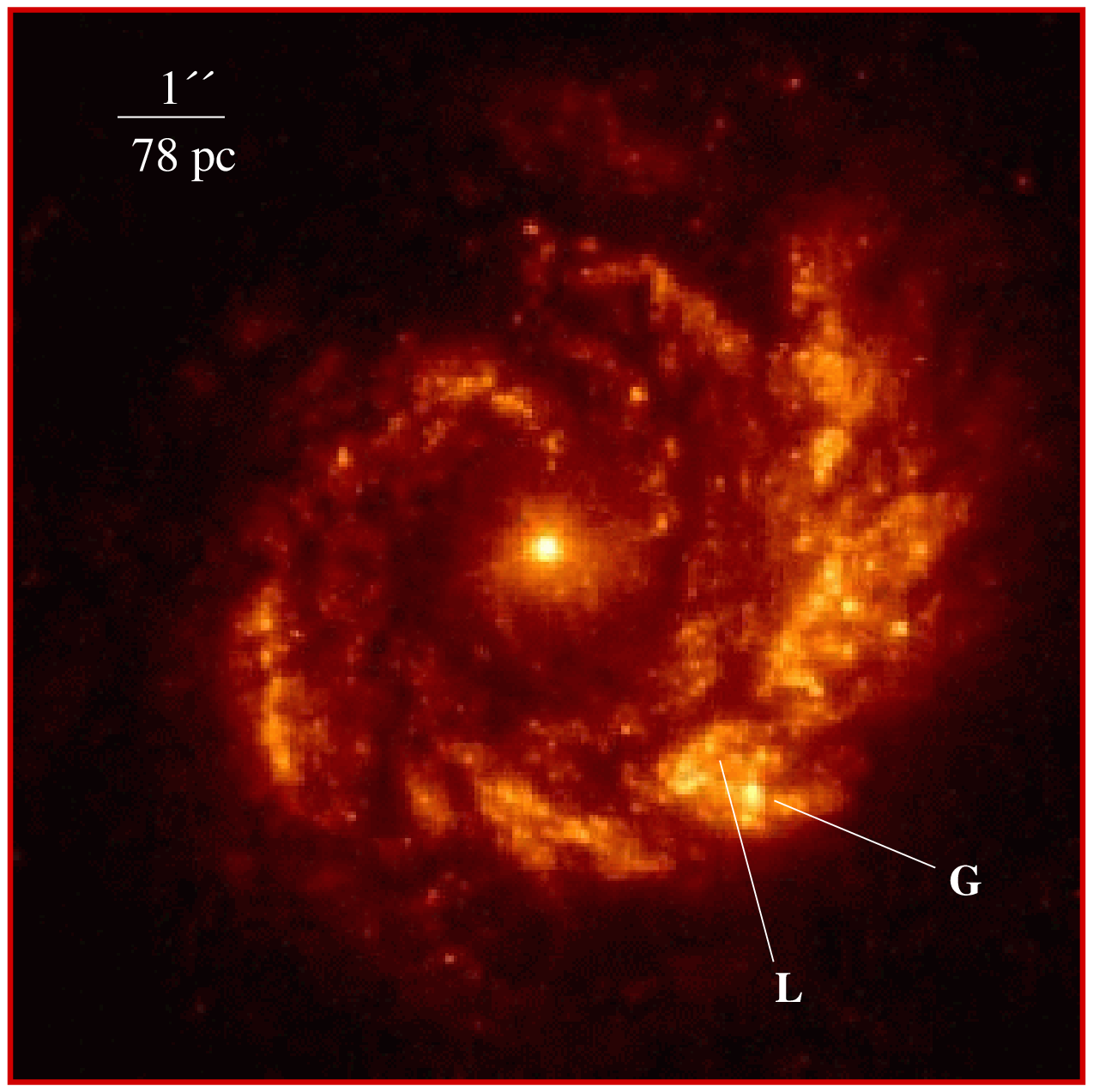,width=17cm,angle=270}} 

\figcaption{STIS F25QTZ ultraviolet image of NGC 4303 nucleus and the surrounding
            star-forming spiral structure, already detected in a previous WFPC2 image 
            (Colina et al. 1997b). The positions of clusters G and L detected in our
            long slit STIS spectrum are indicated for reference. North is up and east 
            to the left.}
\end{figure}


\begin{figure} 

\centerline{\psfig{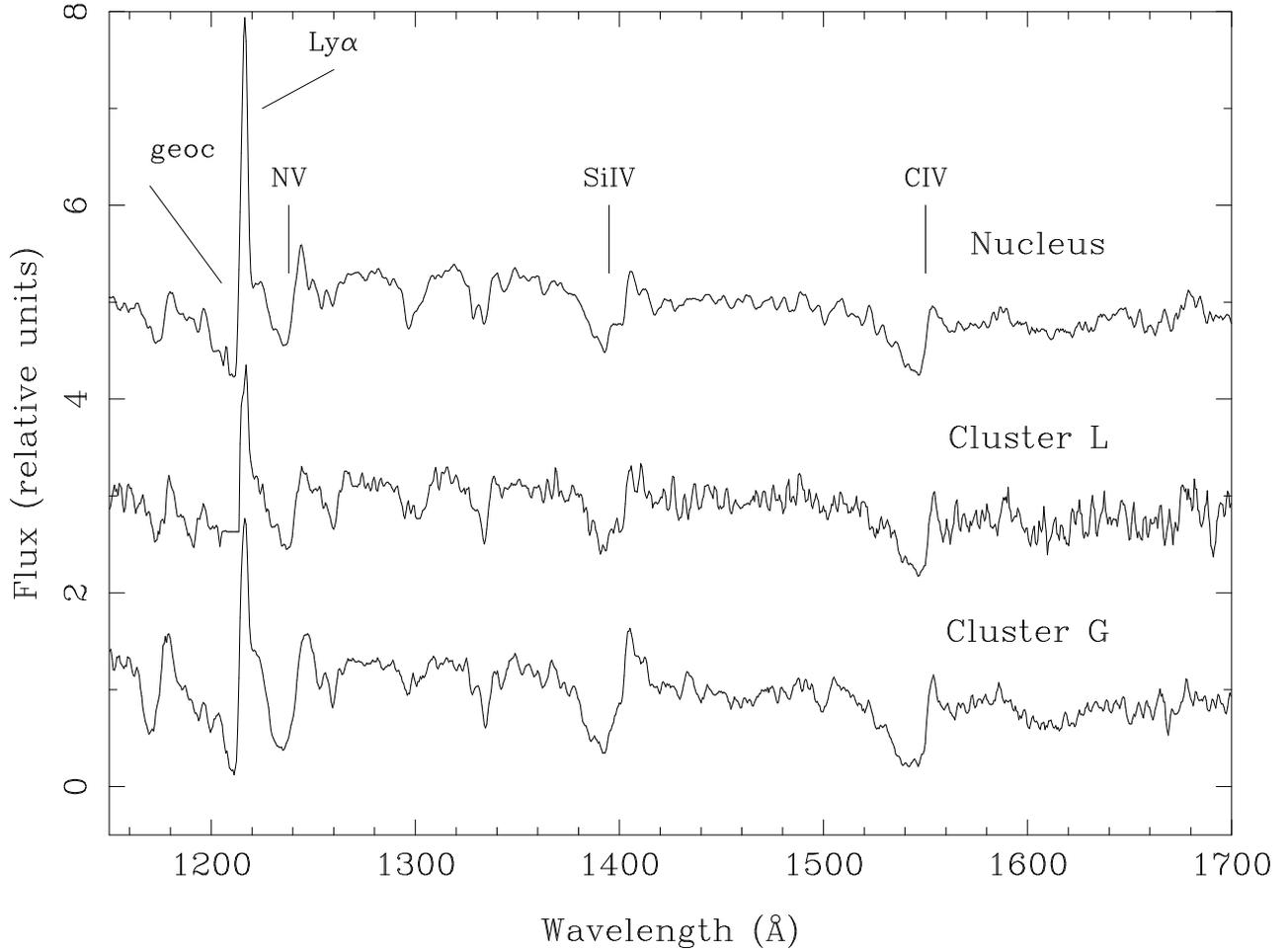}} 

\figcaption{Rest-frame STIS UV spectra of the nucleus and circumnuclear clusters G and L.
            The three spectra show the characteristic 
            broad P Cygni lines (\ion{N}{5} 1240\AA, \ion{Si}{4} 1400\AA, 
            \ion{C}{4} 1550\AA) produced 
            by the winds of massive young stars. In addition, the three spectra show a 
            narrow Ly$\alpha$ line in emission but no trace of other emission lines 
            like \ion{Si}{4} 1400\AA\ and \ion{C}{4} 1550\AA, typical of luminous Seyferts.}

\end{figure} 


\begin{figure} 

\centerline{\psfig{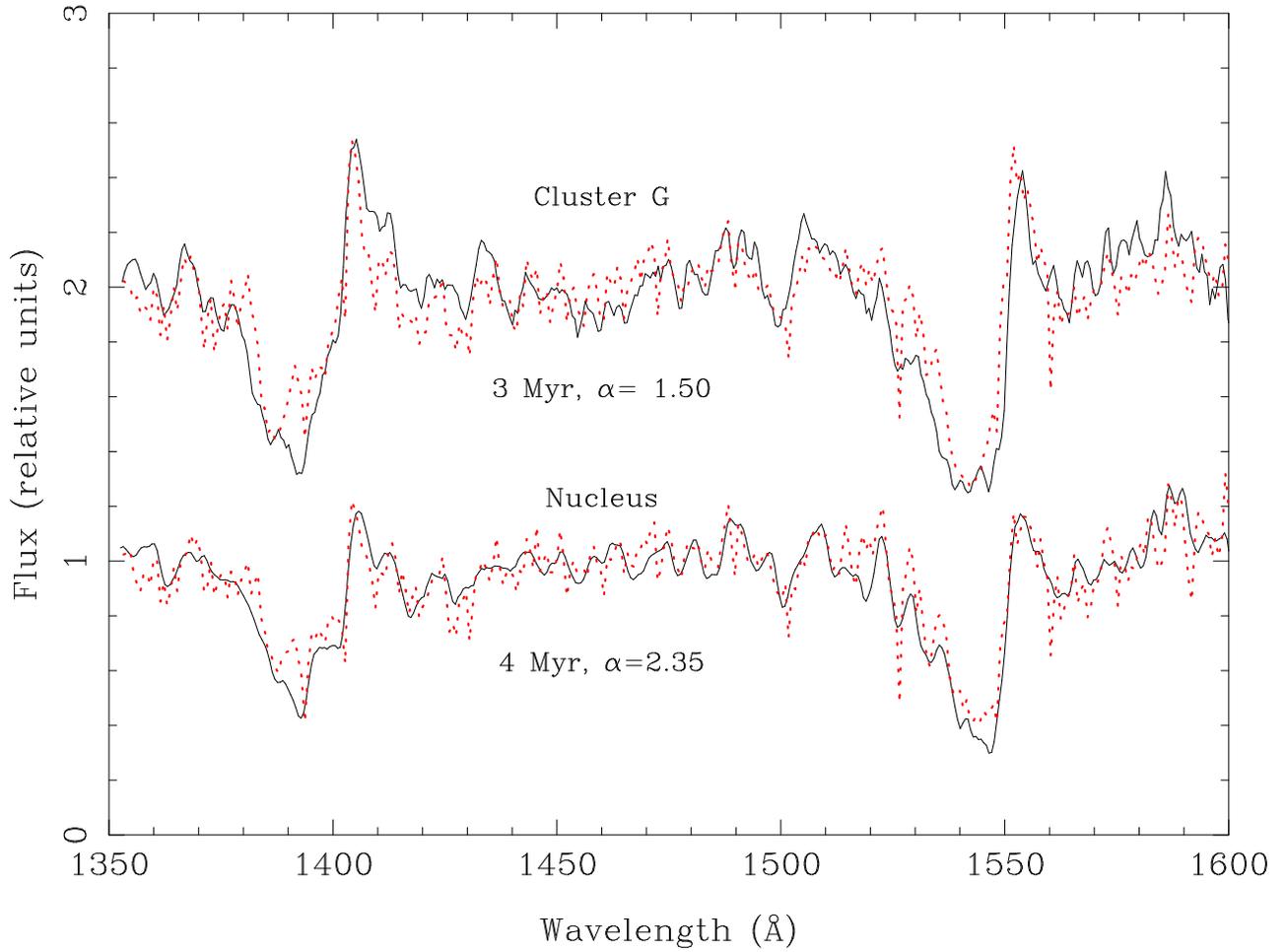}} 

\figcaption{Rest-frame STIS UV spectra of the nucleus and cluster G, centered on the 
            \ion{Si}{4} 1400\AA\ 
            and \ion{C}{4} 1550\AA\ lines, are shown (thin continuous line). The best fitted 
            instantaneous 
            starburst model for the observed spectra are superimposed (dotted lines). The
            age and slope of the IMF for each of the modeled clusters are also indicated.}
\end{figure} 


\begin{figure} 

\centerline{\psfig{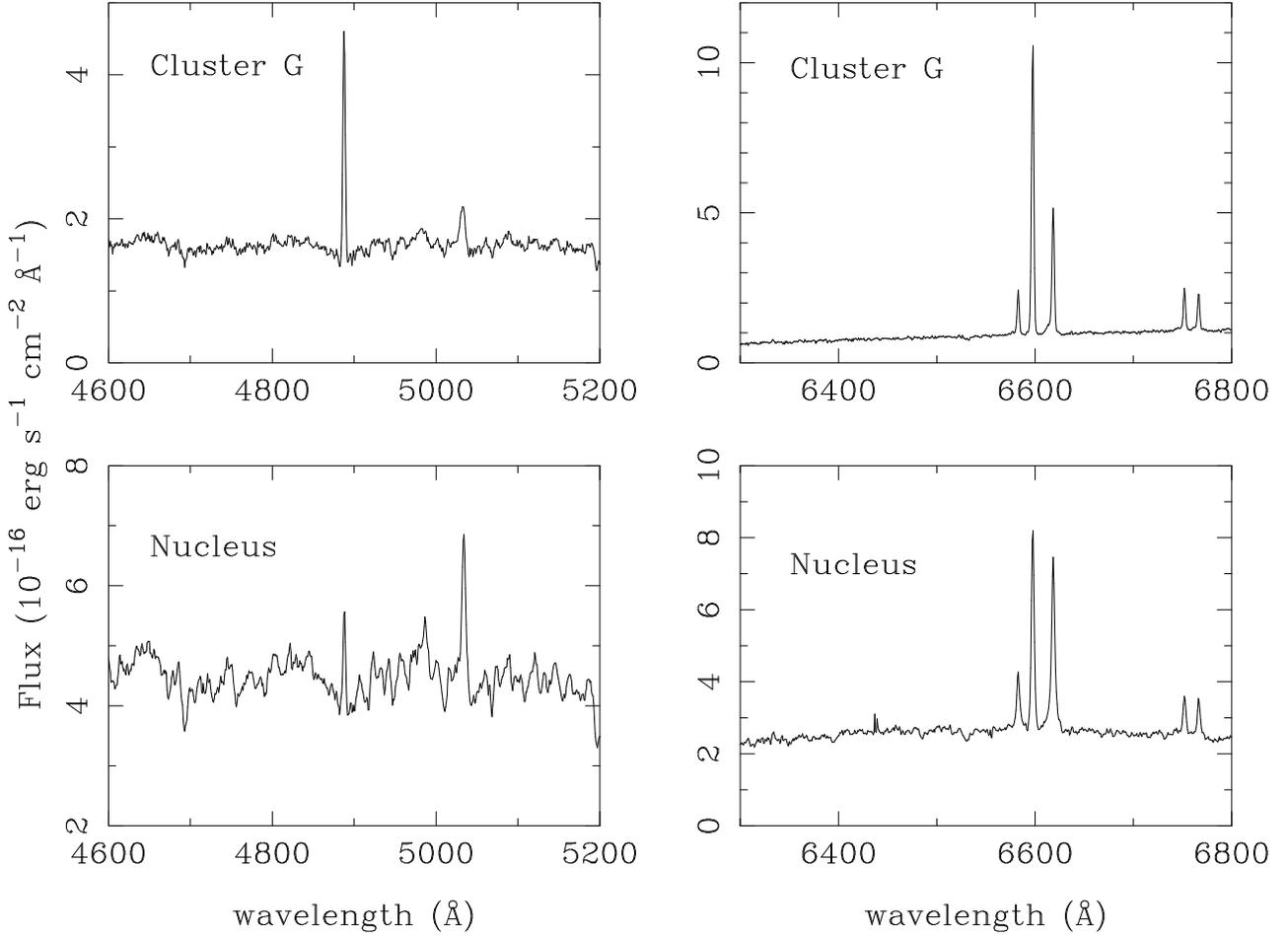}} 

\figcaption{WHT optical observed spectra of the nucleus and cluster G illustrating 
           the different
           excitation conditions in these regions as traced by the changes in the observed
           ratios of the optical emission lines. The left-side plots show the H$\beta$ and
           \ion{[O]}{3} 4959,5007\AA\ part of the spectra while the right-side plots present the 
           H$\alpha$, \ion{[N]}{2} 6548,6584\AA\ and the \ion{[S]}{2} 6717,6731\AA\ lines. 
           The wavelength axis indicates the observed wavelength, not redshift corrected.}

\end{figure} 
     \newpage

\end{document}